\documentclass[conference,a4paper, 10pt]{IEEEtran}
\IEEEoverridecommandlockouts
\usepackage{cite}
\usepackage{amsmath,amssymb,amsfonts}
\usepackage{algorithmic}
\usepackage{graphicx}
\usepackage{textcomp}
\usepackage{comment}
\usepackage{xcolor}
\usepackage{graphicx}
\usepackage{float}
\usepackage{esvect}
\usepackage{color, colortbl}
\usepackage{hhline}
\usepackage{tikz}
\usetikzlibrary{positioning}

\newcommand\copyrighttext{%
  \footnotesize \textcopyright 2012 IEEE. Personal use of this material is permitted. Permission from IEEE must be obtained for all other uses, in any current or future media, including reprinting/republishing this material for advertising or promotional purposes, creating new collective works, for resale or redistribution to servers or lists, or reuse of any copyrighted component of this work in other works. This paper has been accepted for publication in the 2020 Twelfth International Conference on Quality of Multimedia Experience (QoMEX).}
\newcommand\copyrightnotice{%
\begin{tikzpicture}[remember picture,overlay]
\node[anchor=south,yshift=10pt] at (current page.south) {\fbox{\parbox{\dimexpr\textwidth-\fboxsep-\fboxrule\relax}{\copyrighttext}}};
\end{tikzpicture}%
}

\definecolor{Gray}{gray}{0.95}

\def\BibTeX{{\rm B\kern-.05em{\sc i\kern-.025em b}\kern-.08em
    T\kern-.1667em\lower.7ex\hbox{E}\kern-.125emX}}
\begin{document}

\title{Development and Validation of \\ Pictographic Scales for Rapid Assessment\\ of Affective States in Virtual Reality
}

\author{
 \IEEEauthorblockN{Christian Kr\"uger$^1$, Tanja Koji\'{c}$^1$, Luis Meier$^1$, Sebastian M\"oller$^{1,2}$, Jan-Niklas Voigt-Antons$^{1,2}$}
 \IEEEauthorblockA{$^1$Quality and Usability Lab, TU Berlin, Germany\\
 $^2$German Research Center for Artificial Intelligence (DFKI), Berlin, Germany}
}


\maketitle
\copyrightnotice

\begin{abstract}
This paper describes the development and validation of a continuous pictographic scale for self–reported assessment of affective states in virtual environments. The developed tool, called Morph A Mood (MAM), consists of a 3D character whose facial expression can be adjusted with simple controller gestures according to the perceived affective state to capture valence and arousal scores. It was tested against the questionnaires Pick-A-Mood (PAM) and Self-Assessment Manikin (SAM) in an experiment in which the participants (N = 32) watched several one-minute excerpts from music videos of the DEAP database within a virtual environment and assessed their mood after each clip. The experiment showed a high correlation with regard to valence, but only a moderate one with regard to arousal. No statistically significant differences were found between the SAM ratings of this experiment and MAM, but between the valence values of MAM and the DEAP database and between the arousal values of MAM and PAM. In terms of user experience, MAM and PAM hardly differ. Furthermore, the experiment showed that assessments inside virtual environments are significantly faster than with paper-pencil methods, where media devices such as headphones and display goggles must be put on and taken off.
\end{abstract}



\section{Introduction \& Related Work}
Affective states are an important aspect of user experience (UX) \cite{tchatokey2016,norman2004,din2011} and have the advantage to be easily illustratable, not at least because recognizing affective states in daily life is much about the interpretation of visual cues such as mimics, gestures, and postures. This project focuses on the measurement of affective states in order to approach a proof of concept for UX measurement with pictographic scales visualized in 3D. For UX evaluation in virtual reality applications, there are various questionnaires \cite{bowman2002}. Virtual environments (VEs), which are displayed with head-mounted displays (HMDs), differ from "Window-on-the-World" \cite{Milgram1995} systems such as desktop computers or mobile phones among other things by tracking of the user's position and orientation, which allows to display a corresponding egocentric and stereoscopic view of a synthesized virtual space around the user \cite{Ellis1991}. 
The name \textit{Morph A Mood} was selected due to its distinctive feature of continuously interpolatable expressions and as a tribute to the \textit{Pick-A-Mood} (PAM) questionnaire \cite{desmet2016a}, that asks to choose from a set of cartoons the one that best suits one's mood.
For this purpose, a pictographic scale is designed, aiming to be more intuitive to use than a purely text-based scale and more precise than a discrete scale. Disruptive effects can be minimized, which would otherwise be caused by switching between different environmental conditions (i.e., removing the goggles to be able to rate on a screen or paper). It is also assumed that the evaluation process can be accelerated by avoiding the change of environment. Related approaches to pictographic scales for the assessment of affective states are beside PAM, for example, the Self-Assessment Manikin (SAM) \cite{bradley1994} and the affect button \cite{broekens2013}. Furthermore, the Circumplex Model of Affect \cite{russell1980} is a schematic concept of graphically showing the relationship between different states of affect. 
In \cite{Toet2019} a grid of smileys was implemented as an emotional grid to measure emotional responses in VR and utilized 360° videos. The used smileys were implemented as 2D pictures.

\section{Methods}

\subsection{Tool design}
\begin{figure}[htbp]
\includegraphics[width=0.49\textwidth]{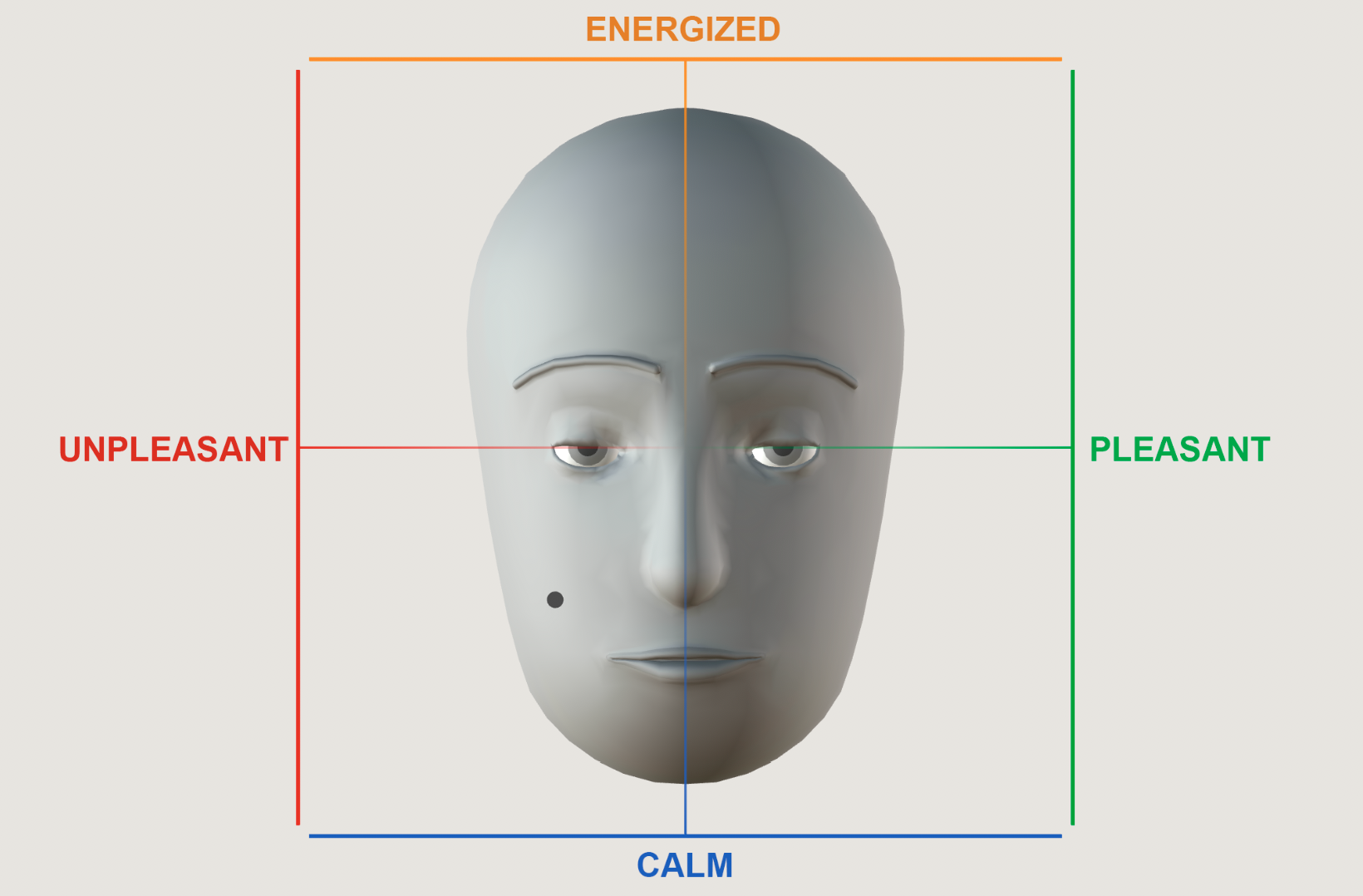}
\caption{Graphical User Interface of MAM in edit mode. The dark dot represents a manually controllable selection cursor. To edit, move the handheld controller and press the trigger button, the colored grid also pops forward as an orientation aid. End editing by releasing the trigger button, which turns the grid gray and moves it to the background to focus on the expression.}
\label{fig:GUI}
\vspace{-1em}
\end{figure}

The graphical user interface (GUI) of the MAM tool consists of a face and a coordinate system with a dot cursor (Fig. \ref{fig:GUI}), layered on top of each other. It was implemented as a Unity-3D asset. The interaction is performed by the movement of a hand-held controller, which triggers changes in cursor position and facial expression. The operation is divided into two modes: a default view mode, which displays the facial expression so that the user can check whether he/she can identify with the representation of the affective state of the figure. And an edit mode that allows the user to modify the expression of the figure until he/she feels that it represents his/her affective state. Editing is not possible in view mode. The separation into the two modes is provided to protect the face from unintentional changes, because the user's hand and thus the controller moves permanently, even slightly.

The character (Fig. \ref{fig:GUI}) is reduced to basic geometric shapes and colored neutral gray to keep it as free as possible from ethnic or gender-specific attributions. It omits ears and hair, because they are not relevant for the interpretation of the expression, as emoticons normally do not have ears. But it has a nose-alike bulge, as it serves as a cue for the expression \textit{irritated} on which it widens as if the nostrils had expanded.

\subsection{Expression Interpolation}
\begin{figure}[htbp]
\centerline{\includegraphics[trim={0.5cm 0cm 0.5cm 0cm}, scale=.69]{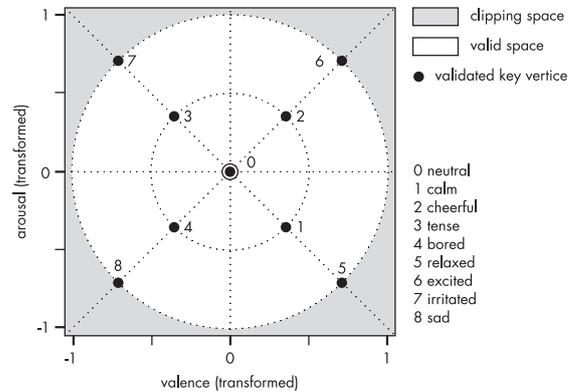}}
\vspace{-0.5em}
\caption{Polar map of expressions. Each vertex represents a constant key expression. For each inserted variable vertex (based on a controlled cursor) an associated expression is interpolated from the nearest neighbour keys.}
\label{fig:polarMap}
\vspace{-0.75em}
\end{figure}

The variety of expressions is possible through interpolating between nine basic key expressions, the same as the 2D model PAM: neutral, calm, relaxed, happy, excited, irritated, tense, sad, bored. The facial features of these expressions were basically derived from the nine cartoons of the Pick-A-Mood questionnaire and from the facial photographs of Ekman and Friesen (Pictures of Facial Affect), Russell \cite{russell1997}, and the facial matrix of the pleasure-arousal space by Grammer and Oberzaucher \cite[pp. 11--12]{grammer2006}. The character was modeled with Blender using \textit{Shape Keys} to parameterize the features. Relevant indications in the cartoons of PAM are gestures. In contrast to facial expressions, these are discrete and not interpolatable. MAM is not displaying gestures. The interpolation is executed on a specific feature vector (FV) per key expression. The FV contains parameters for controlling the curvature and opening of the mouth, the bending of the eyebrows, and the closing of the eyes. It is assumed that the key expressions of MAM have approximately the same average valence and arousal (VA) scores as those of PAM. Therefore VA scores are computed by interpolation between the mean VA scores of the PAM cartoons, which are mapped bijectively to the MAM key expressions (for each expression of MAM exists only one expression of PAM and vice versa). Computationally, each key expression is represented as a vertex on a polar-coordinate map (PCM), that approximates an equidistant, double ring-shaped group of vertices (Fig. \ref{fig:polarMap}). 

\begin{figure}[htbp]
\centerline{\includegraphics[trim={0cm 1.5cm 0cm 0.1cm},clip,scale=.69]{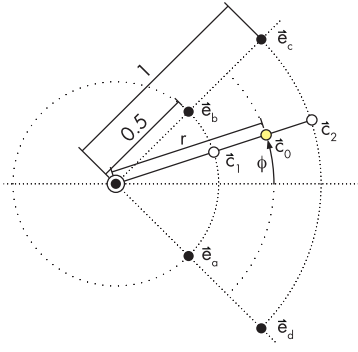}}
\caption{Expression interpolation on controller defined cursor $\Vec{c}_0$ based on the vertices of the nearest key expressions $\Vec{e}_a$, $\Vec{e}_b$, $\Vec{e}_c$ and $\Vec{e}_d$.}
\label{fig:interpolScheme}
\vspace{-1em}
\end{figure}

Similar to the Circumplex Modell \cite{russell2003}, the angle encodes the affective state. MAM uses the radius as a kind of intensity, for example, the state calm has a smaller radius than relaxed, but both have similar angles. The interpolation with the PCM works field related, which means that only the nearest vertices (expressions) of the field containing the cursor are considered. The map based on the VA scores of the PAM expressions does not cover the whole VA space, that was originally provided during the validation of the PAM cartoons \cite[p. 263]{desmet2016a}. The convex hull, defined by the VA values of the vertices, which are used for linear interpolation, is smaller than this VA space. Therefore, only values inside the convex hull can be selected with MAM. Fig. \ref{fig:interpolScheme} shows an exemplary angular segment of the PCM. Each field of the outer ring in the PCM is defined by four vertices, the fields within the inner ring by three vertices each. When computing a new expression for the adjusted cursor $\Vec{c}_0$, the first step is to interpolate two feature vectors $\vv{FV}(\Vec{c}_1)$ and $\vv{FV}(\Vec{c}_2)$ for the temporary cursors $\Vec{c}_1$ and $\Vec{c}_2$ along the angular direction. $\Vec{e}_n$, $n \in \{a,b,c,d\}$, denotes the closest key expressions of the field containing the cursor.

\begin{equation} \label{eq:1}
\vv{FV}(\Vec{c}_1)=(1-\Theta) \cdot \vv{FV}(\Vec{e}_a) + \Theta \cdot \vv{FV}(\Vec{e}_b) 
\end{equation}
\vspace{-2.5em}

\begin{equation} \label{eq:2}
\vv{FV}(\Vec{c}_2)=(1-\Theta) \cdot \vv{FV}(\Vec{e}_d) + \Theta \cdot \vv{FV}(\Vec{e}_c) 
\end{equation}
\vspace{-2em}

\begin{equation} \label{eq:3}
\Theta= \frac{\phi(\Vec{c}_0)-\phi(\Vec{e}_a)}{\phi(\Vec{e}_b)-\phi(\Vec{e}_a)} \text{, $\phi$ : angle, $\Theta$ : angular ratio} 
\end{equation}

In a second step, the searched feature vector $\vv{FV}(\Vec{c}_0)$ of the cursor $\Vec{c}_0$ is computed by interpolation between $\vv{FV}(\Vec{c}_1)$ and $\vv{FV}(\Vec{c}_2)$ along the radial direction.

\vspace{-1em}
\begin{equation} \label{eq:4}
\vv{FV}(\Vec{c}_0)=(1-R(\Vec{c}_0)) \cdot \vv{FV}(\Vec{c}_1) + R(\Vec{c}_0) \cdot \vv{FV}(\Vec{c}_2) 
\end{equation}


For interpolation, the radius value $r$ of the cursor $\Vec{c}_0$ should be in the range from 0 to 1. Due to the inner-outer ring partition, however, it ranges either from 0 to 0.5 (inner ring) or from 0.5 to 1 (outer ring). Therefore, the radius $r(\Vec{c})$ of cursor $\Vec{c}$ is recomputed with the function $R(\Vec{c})$.

\begin{equation} \label{eq:6}
R(\Vec{c})=\begin{cases}
    2 \times r(\Vec{c}) &\text{, $r(\Vec{c}) \leq 0.5$}\\
    2 \times (r(\Vec{c})-0.5) &\text{, $r(\Vec{c}) > 0.5$}
    \end{cases}
\end{equation}

\vspace{-0.75em}

\subsection{Pretest}
The key expressions of MAM were validated performing an online test with a specially programmed web interface, on which the subjects were shown the MAM expressions one after the other in random order next to a matrix of eight PAM expressions. They were asked to select from the PAM expressions the one corresponding to the MAM expression. The pretest with 55 participants (29 male, 22 female, one others, three preferred not to say), aged between 18 and 74 years (M=37.87, SD=15.16) is divided in four cycles and finally achieved accuracies good enough to use MAM in the main study: 93.55\% for \textit{lively}, 74.19\% for \textit{annoyed}, 67.74\% for \textit{tense}, 87.1\% for \textit{happy}, 90.32\% for \textit{relaxed}, 64.52\% for \textit{bored}, 83.87\% for \textit{sad} and 80.65\% for \textit{calm}. The accuracy describes to what extent the participants chose the corresponding cartoon of PAM.

\subsection{Procedure}
In order to validate MAM, it was tested whether it is possible to measure agreement on the valence and arousal scores of MAM with its already validated 2D counterparts PAM and SAM. Additionally, the UX of the questionnaire GUIs of PAM and MAM was tested using the Questionnaire for the Subjective Consequences of Intuitive Use (QUESI) \cite{naumann2010} and AttrakDiff \cite{Diefenbach2010}, to evaluate how the perceived interaction with the continuously scaled pictographic VR interface differs from the one of the discretely scaled pictographic paper interface. Each subject performed the experiment under two environmental conditions (within-subjects design), i.e., MAM and a VR app of SAM (SAM\textsubscript{VR}) were provided inside the VE, PAM and SAM (SAM\textsubscript{PP}) were provided as paper-pencil versions outside VE. The result of inside VR ratings (VRR) was compared to the result of paper-pencil ratings (PPR). For time measurement, time stamps were automatically generated in the VR applications. On paper-pencil runs, the supervisor manually set timestamps after removing the HMD, after each checkmark and after reattaching the HMD and controller by pressing specific buttons in a specially designed web app.

Using SAM\textsubscript{PP} and SAM\textsubscript{VR} on the same set of stimuli serves to check whether the same rating method measures reliably in different media conditions (in VR vs. on paper). It is assumed that if there are no significant differences between SAM\textsubscript{PP} and SAM\textsubscript{VR}, the medium has no effect, and the VR SAM ratings can also be considered without regard to the different media types. Using two different environmental rating conditions with each having two different rating methods leads to eight cases of the sequential order participants run the different rating methods. The within-subjects design of the experiment takes into account sequential effects and fully balances the different order of conditions between the 32 subjects, so each possible sequential order is performed four times. The three questionnaires MAM, PAM, and SAM, are based on different scales. In order to compare the results from MAM, PAM, and SAM, the values from SAM are scaled from a 9-point range to a 5-point range used to validate the PAM cartoons. Since MAM is based on PAM cartoons and interpolates its assigned VA values, MAM scores are ranging between 1 and 5.
The experiment uses music video stimuli of different genres from the DEAP database \cite{koelstra2012}, which provides valence (V) and arousal (A) ratings for each. The video set covers all quadrants of the VA space. From the 40 videos, a subset was created of those clips that are still available on youtube.com and have sufficient quality of video and audio (assessed by visual inspection). Each video has associated VA scores resulting in a map of videos in VA space. In total, 16 clips of the subset were selected for the validation experiment, using the same excerpts of videos as in the DEAP experiment. The distribution was achieved by including clips from each quadrant of the VA space. After pre-selecting all those clips that had sufficient video quality, i.e., a width of at least 360 pixels and no image glitches, the two clips with the greatest and the smallest distance in VA space to the coordinate origin were selected for each quadrant. It is assumed that the distance corresponds to the intensity of the affect and that in this way, there is one strong and one weak stimulus per quadrant. A clip was also selected for each quadrant, where the valence value is approximately as large as the arousal value. This was calculated by selecting the clip with the smallest deviation of the $\frac{V}{A}$ ratio to the value 1. In addition, four further clips were selected, each with a value that is approximately neutral, either valence or arousal. This was also determined by the $\frac{V}{A}$ ratio. If the V value approaches 0 (the assumed neutral value), the $\frac{V}{A}$ ratio also approaches zero. If the A value goes towards 0, the $\frac{V}{A}$ ratio goes towards infinity. If a stimulus ranks first for two categories, e.g., lowest distance and balanced VA values, it will only be used as a representative of one class, and in the set of the other class, the second-placed candidate will be chosen.
The experiment was conducted using the head-mounted display Oculus Go. For the stimuli display in VR, a floating, slightly curved virtual screen in a dark environment was set up, the width of which adapted to the different ratios. For the audio presentation, a binaural synthesis was used to provide spatial sound. The video clips have different audio volumes and were leveled to a common value of about -12 dB, based on the two videos with the lowest sound level (Song 2 by Blur, Love Story by Tyler Swift). At the Oculus Go device, the volume was adjusted to 11 of the 15 units. Due to the limited frequency spectrum of the integrated speakers of the Oculus Go, headphones (Sennheiser HD 485) were used.
Unfortunately, the video player plug-in of the Unity game engine has difficulty to play high-resolution videos synchronously with the audio signal. The higher the resolution, the stronger and more noticeable the asynchronous signal shift. This becomes particularly clear in\enlargethispage{1\baselineskip} close portrait views. A trade-off between video resolution and delay was approached by downsampling to about 360 x 480 pixels (varying depending on the ratio), whereby the asynchrony was kept to a minimum.

\subsection{Participants}\label{SCM}
The 32 participants (15 male, 17 female), were aged between 19 and 51 years (M=30.38, SD=6.2). Participants received a compensation of 15 €. None of the participants had red-green blindness, but two had blindness on one of the eyes resulting in an inability to see stereoscopically.

\section{Results}
\subsection{Assessment of Affective States}
\label{results_affective_state}
\begin{figure}[h]
\centerline{\includegraphics[trim={0 0.35cm 0 0.9cm},clip,scale=.8]{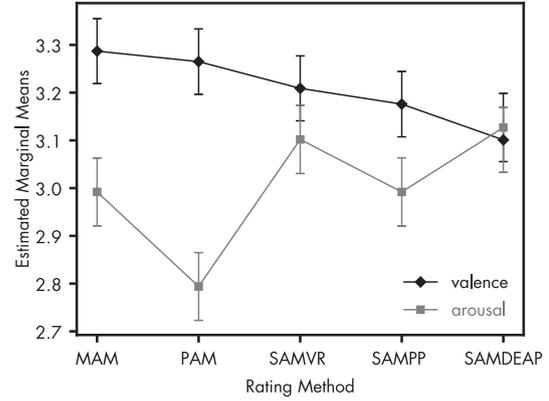}}
\caption{Estimated Marginal Means of valence and arousal scores per rating method. The methods MAM, PAM, SAM\textsubscript{VR}, SAM\textsubscript{PP} were used to assess affective states during this study, SAM\textsubscript{DEAP} was assessed during the DEAP experiment \cite{koelstra2012} and included here for comparison. In both studies participants watched music videos and assessed their affective state after each clip.}
\label{fig:emm}
\vspace{-1em}
\end{figure}

A two-way MANOVA was run with the dependent variables \textit{valence} and \textit{arousal} and with the independent variables \textit{rating method} and \textit{video stimulus}. The dataset, including VA scores of this experiment and the DEAP experiment, did not show in all cases, a linear relationship between the dependent variables as assessed by scatterplots. There is no evidence of multicollinearity, as assessed by Pearson correlation ($|r| < 0.9$), but several univariate outliers were found by inspection of boxplots, and ten multivariate outliers by Mahalanobis distance ($p > 0.001$). 
However, the MANOVA procedure is conducted as it is considered to be robust against violations of the assumption of normal distribution according to Bray \& Maxwell \cite[p. 33]{bray1985} and Weinfurt \cite{weinfurt1995} as well as the violation of the assumption of homogeneity of covariance matrices as long as the sample size in each cell is similar \cite[ pp. 292--294]{tabachnick2014}, which is true in case of this dataset. Due to the violated assumptions Pillai's Trace is used as recommended by Olson \cite{olson1976}. There was a statistically significant interaction effect between \textit{rating method} and \textit{video stimulus} on the combined dependent variables, F(120, 4960) = 1.829, p $<$ 0.001, Pillai's Trace = 0.085, partial $\eta^2$ = 0.42. This was shown also in follow up univariate two-way ANOVAs for valence score, F(60, 2480) = 1.991, p $<$ 0.001, partial $\eta^2$ = 0.046, as well as for the arousal score, F(60, 2480) = 1.584, p = 0.003, partial $\eta^2$ = 0.037. Simple main effects analysis revealed statistically significant difference between the rating methods for both scores. Valence: F(4, 2480) = 4.533, p = 0.001, partial $\eta^2$ = 0.007; Arousal: F(4, 2480) = 12.980, p $<$ 0.001, partial $\eta^2$ = 0.021. 
Further significant differences are shown in table \ref{VAmeanDiff}, estimated marginal means in Fig. \ref{fig:emm}.

\begin{table}[h]
\caption{Absolute mean differences of Valence and Arousal scores}
\begin{center}
\begin{tabular}{c|c|c|c|c|c}
\hhline{|~|-|-|-|-|-|}
\multicolumn{1}{l|}{} &
\multicolumn{5}{c|}{{\cellcolor{Gray}}\textbf{Arousal}} \\ \hhline{|~|-|-|-|-|-|} 
\multicolumn{1}{l|}{} & \textbf{MAM} & \textbf{PAM} & \textbf{SAM\textsubscript{VR}} & \textbf{SAM\textsubscript{PP}}        & \multicolumn{1}{c|}{\textbf{SAM\textsubscript{DEAP}}} \\ \hhline{|-|-|-|-|-|-|}
\multicolumn{1}{|c|}{\textbf{MAM}}     & \multicolumn{1}{l|}{} & {\cellcolor{Gray}}\textbf{0.198*}       & {\cellcolor{Gray}}0.11                  & {\cellcolor{Gray}}$<$ 0.001               & \multicolumn{1}{c|}{{\cellcolor{Gray}}0.135}            \\ \hhline{|-|-|-|-|-|-|}
\multicolumn{1}{|c|}{\textbf{PAM}}     & 0.023                 & \multicolumn{1}{l|}{} & {\cellcolor{Gray}}\textbf{0.308*}       & {\cellcolor{Gray}}\textbf{0.198*}       & \multicolumn{1}{c|}{{\cellcolor{Gray}}\textbf{0.333*}}  \\ \hhline{|-|-|-|-|-|-|}
\multicolumn{1}{|c|}{\textbf{SAM\textsubscript{VR}}}   & 0.078                 & 0.056                 & \multicolumn{1}{l|}{} & {\cellcolor{Gray}}0.109                 & \multicolumn{1}{c|}{{\cellcolor{Gray}}0.025}            \\ \hhline{|-|-|-|-|-|-|}
\multicolumn{1}{|c|}{\textbf{SAM\textsubscript{PP}}}   & 0.111                 & 0.088                 & 0.032                 & \multicolumn{1}{l|}{} & \multicolumn{1}{c|}{{\cellcolor{Gray}}0.134}            \\ \hhline{|-|-|-|-|-|-|}
\multicolumn{1}{|c|}{\textbf{SAM\textsubscript{DEAP}}} & \textbf{0.186*}       & \textbf{0.164*}       & 0.108                 & 0.075                 & \multicolumn{1}{l}{}                  \\ \hhline{|-|-|-|-|-|~|}
\multicolumn{5}{|c|}{\textbf{Valence}}                                                                                                 & \multicolumn{1}{l}{}                  \\ \hhline{|-|-|-|-|-|~|}
\end{tabular}
\end{center}
{$^{\mathrm{*}}$ significant at the 0.05 level. For complete test values see section \ref{results_affective_state}. The methods MAM, PAM, SAM\textsubscript{VR}, SAM\textsubscript{PP} were used to assess affective states during this experiment. SAM\textsubscript{DEAP} was taken from the DEAP database \cite{koelstra2012}.}
\label{VAmeanDiff}
\vspace{-1em}
\end{table}

In order to assess the degree that the rating methods used during this experiment (MAM, PAM, SAM\textsubscript{VR}, SAM\textsubscript{PP}) provided agreement in their ratings across subjects, the inter-rater reliability (IRR) was assessed using a two-way mixed, absolute agreement, average-measures intra-class correlation (ICC) \cite{mcgraw1996,hallgren2012}. The average resulting ICCs regarding the 16 stimuli were of excellent reliability \cite{cicchetti1994} for the valence score, total average ICC = 0.859, p $<$ 0.05, and of good reliability \cite{cicchetti1994} for the arousal score, total average ICC = 0.628, p $<$ 0.05, indicating that the rating methods had a moderate to high degree of agreement and suggesting that at least valence was rated similarly across the rating methods.

In summary, there are no statistically significant differences in the measurement of valence and arousal between SAM\textsubscript{VR}, SAM\textsubscript{PP}, and MAM, but the values of these three methods differ from those obtained with PAM. Compared to the values of the DEAP database captured with SAM, statistically significant differences exist only for PAM and MAM, and for MAM only with regard to the valence values.

\subsection{User Experience of MAM vs. PAM}
To assess the user experience, a one-way MANOVA was run with eight dependent variables (five QUESI scores and three AttrakDiff scores) and the independent variable \textit{expression based rating method} (levels: MAM, PAM). There was not in all cases of the dataset a linear relationship between the dependent variables as assessed by scatterplots, and no evidence of multicollinearity, as assessed by Pearson correlation (MAM: r = 0.418, p = 0.017; PAM: r = 0.393, p = 0.026). There were several univariate outliers in the data, as assessed by inspection of boxplots, but no multivariate outliers, as assessed by Mahalanobis distance (p $>$ 0.001). All QUESI scores are not normally distributed, AttrakDiff scores are normally distributed except in case of the \textit{Pragmatic Quality} scores regarding PAM, as assessed by Shapiro-Wilk's test (p $>$ 0.05). 
The MANOVA is considered to be robust against violations of the assumption of normal distribution according to Bray \& Maxwell \cite[p. 33]{bray1985} and Weinfurt \cite{weinfurt1995}. 

\begin{figure}[hb]
\centerline{\includegraphics[trim={0 0.35cm 0 0.8cm},clip,scale=.8]{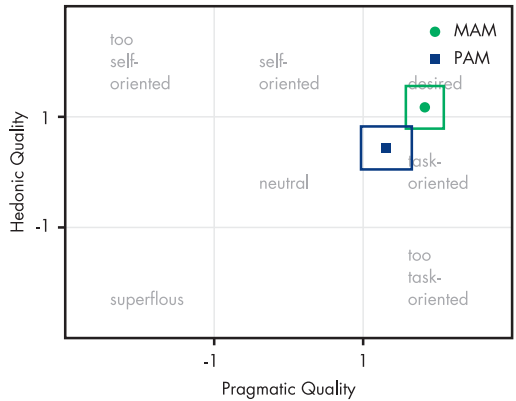}}
\vspace{-0.5em}
\caption{AttrakDiff Portfolio view showing the mean Pragmatic Quality and Hedonic Quality scores. Each box represents the confidence intervals of the corresponding mean value pair represented as dot.}
\label{fig:attrakdiff}
\vspace{-1em}
\end{figure}

There was homogeneity of variance-covariances matrices, as assessed by Box's M test (p = 0.672), and homogeneity of variances in case of the QUESI scores \textit{Subjective Mental Workload, Perceived Effort of Learning, Perceived Error Rate} and the AttrakDiff scores, as assessed by Levene's Test (p $>$ 0.05). \textit{Perceived Achievement of Goals} (PAG) and \textit{Familiarity} scores have statistically significant differences in variance. However, the MANOVA procedure is conducted as it is considered to be robust against such violations as long as the sample size in each cell is similar \cite[pp. 292--294]{tabachnick2014}. Due to the partially violated assumptions Pillai's Trace is used as recommended by Olson \cite{olson1976}. There was a statistically significant difference between the rating methods MAM and PAM on the combined dependent variables, F(8, 55) = 3.966, p $<$ 0.001, Pillai's Trace = 0.366, partial $\eta^2$ = 0.366. Follow up univariate one-way ANOVAs showed statistically significant differences between the rating methods MAM and PAM for the QUESI PAG score, F(1, 62) = 6.102, p = 0.016, partial $\eta^2$ = 0.09, for the AttrakDiff \textit{Pragmatic Quality} (PQ) score, F(1, 62)=5.46, p=0.023, partial $\eta^2$=0.081, as well as for the AttrakDiff \textit{Hedonic Quality} (HQ) score, F(1, 62) = 6.704, p = 0.012, partial $\eta^2$=0.098. MAM (M=3, SD = 0.179) scored 0.65 higher for the PAG score than PAM (M = 2.375, SD = 0.179). 
The differences in PQ and HQ are shown in Fig. \ref{fig:attrakdiff}.

In summary, although there are statistically significant differences between MAM and PAM with respect to three UX scores, these differences are minimal.

\subsection{Duration per Rating Mode}
To determine the effect of the two rating modes (VRR vs. PPR) over the different video stimuli on the duration of the assessment, a two-way repeated-measures ANOVA was performed. Due to the violated assumptions of normality (Shapiro-Wilk) and sphericity (Mauchly; $\chi^2$(2) = 1782.781, p $<$ 0.01) Greenhouse-Geisser correction has been applied as suggested by Maxwell and Delaney \cite{maxwell2004}. No statistically significant interaction between rating mode and stimulus on duration of assessment was found, F(1.231, 38.161) = 0.869, p = 0.371, partial $\eta^2$ = 0.028. The main effect of rating mode showed a statistically significant difference in the duration of assessment, F(1, 31) = 17.275, p $<$ 0.001, partial $\eta^2$ = 0.358 and pairwise comparison indicates, that VRR mode (M = 24.744 s, SD = 2.127) is 96.631 s, 95\% CI [-144.048, -49.214], faster than PPR mode (M = 121.375 s, SD = 23.346).

In summary, evaluations within VE are statistically significantly faster than evaluations outside VE on paper.

\subsection{Duration per Rating Method}
To determine the effect of this study's rating methods (MAM, PAM, SAM\textsubscript{VR}, SAM\textsubscript{PP}) over different video stimuli on the duration of the assessment, a two-way repeated-measures ANOVA was run, which did not show any statistically significant interaction between rating method and stimulus on the duration of the assessment, F(1.745, 54.088) = 1.547, p = 0.224, partial $\eta^2$ = 0.048. But the main effect of the rating method showed a statistically significant difference in the duration of assessment, F(1.489, 46.168) = 7.462, p = 0.004, partial $\eta^2$ = 0.194 and as such, pairwise comparisons showed statistically significant differences, i.e. on average PAM is 6.314 s, 95\% CI [4.604, 8.025], faster than SAM\textsubscript{PP}, and SAM\textsubscript{VR} is 4.082 s, 95\% CI[1.679, 6.485], faster than SAM\textsubscript{PP}. But MAM was not significantly faster than the other methods. Due to violated assumptions of normality and sphericity (Mauchly failed), Greenhouse-Geisser correction has been applied \cite{maxwell2004}.

In summary, there are no statistically significant differences in processing time between MAM and the other methods, but among them, PAM and SAM\textsubscript{VR} are faster than SAM\textsubscript{PP}.


\section{Discussion \& Conclusion}
\subsection{Assessment of Affective States}
Compared to PAM, MAM measures valence with approximately the same precision, since no statistically significant differences were found in the valence values and also high ICC values regarding the valence measure suggest a high degree of agreement between the rating methods. But there are statistically significant differences in the mean of the arousal measure between PAM and all other rating methods. Even if the average ICC\textsubscript{arousal} value (0.628) indicates a good agreement according to Cicchetti \cite{cicchetti1994}, which suggests a certain level of agreement on the measures, nearly half of the lower bounds of the confidence intervals of ICCs respective the arousal score are in the poor range (ICC $<$ 0.4). The lower agreement in the assessment of arousal may be related to the statistically significant differences between the arousal values of PAM and those of MAM, SAM\textsubscript{VR}, SAM\textsubscript{PP}, and SAM\textsubscript{DEAP}. The differences are presumably due to the fact that PAM, unlike the other methods, has a discrete scale that also binds each arousal value firmly to a valence value. There are no statistically significant differences between MAM and the SAM versions of this study regarding the two dependent variables, valence and arousal. There are also no statistically significant differences between MAM and SAM\textsubscript{DEAP} with regard to the arousal score, but with regard to valence. The SAM rating methods – SAM\textsubscript{VR}, SAM\textsubscript{PP}, and SAM\textsubscript{DEAP} – seem to be comparable, as no statistically significant differences (p $>$ 0.05) were found. Given the approximately same measurement results between MAM and the SAM versions and the differences between PAM and all other methods regarding the arousal score, it can be concluded that the continuously scaled tools measured differently than the discretely scaled PAM tool. It remains questionable whether the agreement between MAM and the SAM versions is principally based on the continuous scaling or whether it is induced by MAM's additional coordinate system, which is graphically similar to the SAM bar scale. The question of whether the agreement would still exist even if MAM would not provide the additional coordinate system and only consists of the continuously morphable facial expression cannot be answered with this study.

\subsection{UX}
The differences between MAM and PAM regarding the UX criteria are small in number and impact. The total QUESI score of PAM is only 0.3 (of the 5-point scale) higher than that of MAM, which is why no great differences can be assumed with regard to intuitivity. The Pragmatic Quality score of MAM is 0.516 higher, and the Hedonic Quality score of MAM 0.734 higher than those of PAM, but both tools have very good values according to the AttrakDiff Portfolio view. In the personal feedback, the interactivity of MAM was highlighted in particular. The fact that users are allowed to set and adjust something themselves seems to have an effect. It should be clarified to what extent the increased identification comes solely from the modification in terms of appropriation or from the suitability of the personally generated expression. It is reasonable to assume that PAM would have achieved different UX with an additional intensity scale per cartoon element and an implementation inside VR.

\subsection{This Study vs. the DEAP experiment}
The media condition (VR vs. screen) under which the chosen stimuli were presented does not seem to have a strong influence on the measurement of the affective state since statistically significant differences between VA values obtained with SAM in this experiment (both with VRR and PPR) and the SAM values from the DEAP experiment were not found.

\subsection{Inside VR ratings vs. paper-pencil ratings}
The differences in performance between VRR and PPR are significant. On average, VRR is up to 490\% faster than PPR, since no time is required to remove and set up the HMD during VRR runs. However, the advantage of VRR over PPR does not mean that 3D pictographic scales, in this case, MAM, are faster. The speed advantage refers mostly to the change of the media environment, not to the evaluation model since the average rating times per rating method are quite similar.


\bibliographystyle{IEEEtran}
\bibliography{references}

\vspace{12pt}

\end{document}